\documentclass[12pt,showpacs,preprintnumbers,amsmath,amssymb]{revtex4-1}
\usepackage{graphicx}
\usepackage{fancyhdr}
\usepackage{pslatex}
\usepackage{amsfonts}
\usepackage[mathcal]{eucal}
\usepackage{latexsym}
\begin{document}

\newcommand{\pderiv}[2]{\frac{\partial #1}{\partial #2}}
\newcommand{\deriv}[2]{\frac{d #1}{d #2}}

\title{Entropic considerations on the Two-Capacitor Problem}

\vskip \baselineskip

\author{V.O.M. Lara}

\author{A. P. Lima}

\author{A. Costa}


\affiliation{Instituto de F\'{\i}sica - Universidade Federal Fluminense \\
Av. Litor\^anea s/n \\
24210-340 \hspace{0.5cm} Niter\'oi - RJ \hspace{0.5cm} Brazil}

\date{\today}

\begin{abstract}
In the present work we study the well-known Two Capacitor Problem from a new perspective. Although this problem has been thoroughly investigated, as far as we know there are no studies of the thermodynamic aspects of the discharge process. We use the Free Electron Gas Model to describe the electrons' energy levels in both capacitors in the low temperature regime. We assume that the capacitors and the resistor can exchange energy freely with a heat reservoir. We assume that the resistance is large enough to consider an isothermal heat exchange between the resistor and the heat reservoir. Thereby we obtain a positive entropy variation due to the discharge process, corroborating its irreversibility.
\end{abstract}

\maketitle

\vskip \baselineskip

\section{Introduction}

\begin{figure}[!htb]
\begin{center}
\vspace{0.6cm}
\includegraphics[scale=0.25]{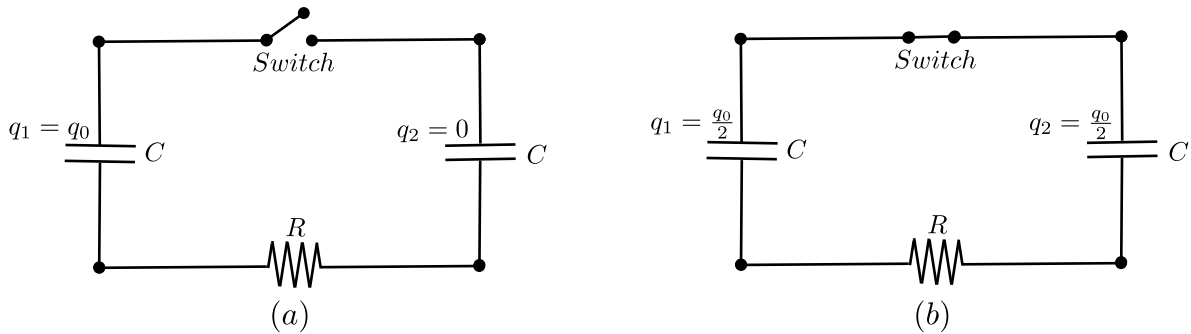}
\end{center}
\caption{Illustration of the circuit described in the text. In $(a)$ the switch is open and there is no current flowing. In $(b)$, we have the final equilibrium situation, in which a sufficiently large time has elapsed for the system to reach equilibrium.}
\label{circuit}
\end{figure}

In order to compute the entropy change $\Delta S$, suppose that the circuit of Fig. (\ref{circuit}) is in thermal contact with a heat reservoir at temperature $T_{0}$. Consider a simple circuit that consists of two identical capacitors, with the same capacitive constant $C$, and one resistor $R$, all connected in series. One of the capacitors (capacitor 1) is initially charged with a charge $q_0$, while the other one has zero charge (capacitor 2). The circuit also has a switch that prevents the flow of current, as shown in figure (\ref{circuit}). 

If we close the switch a current will flow until the two capacitors reach the same potential $V$ (in our case, $V = \frac{q_{0}}{2C}$, since the capacitors have the same capacitive constant $C$). We know intuitively, that the discharging process is highly irreversible: the system reaches spontaneously the equilibrium configuration described above and it will not go back to the initial configuration without external interference. There is a simple analogy between this discharging process and the isothermal expansion of an ideal gas (commonly used to obtain the entropy change in a free expansion). In both cases, there is a positive entropy change for the gas, due to the redistribution of particles. However, here the particles in the play are fermions, thus obeying a different statistical distribution, which is a fundamental difference between then.

There are many previous works which deal with this simple system using a variety of approaches. In \cite{cujav, powell}, the authors focus on energy considerations, while \cite{salih, boykin, choy} discuss specifically the electromagnetic radiation produced by the discharge process. Some of those works neglect the circuit's electrical resistance, but include a self inductance. Some authors consider both resistance and inductance. 

In \cite{heinrich} the author measures some quantities in order to obtain the entropy change of the capacitor charging process. The author shows experimentally that if the charging process can be considered a quasistatic one, the entropy change due to resistor and the heat reservoir adds to zero. In our discharging problem, this limiting case is achieved by a a large resistance $R$. However, this work considers only the entropy variation due to heat exchange between the resistance and the heat reservoir, obtaining $\Delta S = 0$ when the process is sufficiently slow. We will use a simple model to discuss the entropy variation of the whole system. 

It is our goal to calculate the change in the thermodynamical entropy $S$ due to the discharge process. We will show how this can be done using a simple model to describe the electrons in the capacitors, allowing a straightforward evaluation of $\Delta S$. As mentioned before, the discharging process is clearly irreversible, such that any model chosen to describe the discharging process must provide $\Delta S > 0$.

\section{Entropy Calculation}

According to the second law of thermodynamics, any process that occurs in an isolated system must have $\Delta S \geq 0$. So, to compute $\Delta S$, we need the entropy change of all components of our system: both capacitors, the heat reservoir and the resistance. First, we will evaluate $\Delta S$ for the heat reservoir and the resistance.

As mentioned above, we consider that the resistance is large enough, in order that  there is a isothermal heat exchange between the resistor and the heat reservoir. Thus $\Delta S_R=-\Delta S_{HR}$ and the total entropy change due to this exchange is zero. As a consequence the total entropy variation of the discharge process is only due to the charge rearrangement in the two capacitors. To calculate this variation we will use the Free Electron Gas model to describe the electrons in the metal plates of the capacitors. We can consider the plates as three dimensional: although their thicknesses are very small in comparison to the other dimensions, they are indeed macroscopic as far as the motions of electrons are concerned.

The Free Electron Gas Model was the first attempt at a microscopic description of the properties of metals. It was first used by Drude\cite{ashcroft} and later modified by Sommerfeld \cite{sommerfeld,ashcroft} to include the quantum nature of the electrons. The model consists of the assumption that electrons are confined to a box with impenetrable walls. Inside the box the electrons are subject to a uniform potential $V_0$. Although this model is very simple, it describes qualitatively very important features of metals, such as the Wiedemann-Franz Law and the fact that the electronic contribution to the specific heat is proportional to $T$ at low temperatures (as compared to the Fermi temperature) \cite{ashcroft,sommerfeld}.

Since the model assumes that electrons do not interact with each other, the solution to the problem consists of finding the eigenstates of a quantum particle confined to a box and filling up these states according to Fermi-Dirac statistics. In order to obtain the expression for the electronic contribution to the entropy, we will use a well-known result for the specific heat in this model\cite{ashcroft,kittel}),

\begin{equation}
 C_{V} = \frac{N k_{B}\pi^{2}}{2} \frac{k_{B}T}{\epsilon_{F}} , 
\label{sommerfeld_specific_heat}
\end{equation}
\noindent
where $k_{B}$ is the Boltzmann constant and $\epsilon_{F}$ is the Fermi energy which is given by

\begin{equation}
 \epsilon_{F} = \frac{\hbar^{2}k_{F}^{2}}{2m} = \frac{\hbar^{2}}{2m} \left(\frac{3\pi^{2}}{V}\right)^{2/3} N^{2/3} ,
\label{fermi_energy}
\end{equation}
\noindent
where $N$ is the total number of free electrons, $\hbar$  is the Planck constant divided by $2\pi$, $m$ is the electron mass and $V$ is the volume of the macroscopic sample. The expression (\ref{sommerfeld_specific_heat}) results from the Sommerfeld expansion for the specific heat, which is very accurate as long as the temperature is small when compared to the Fermi temperature $T_{F}$. This is easily satisfied at room temperature, at which $T/T_{F} \sim 10^{-2}$).

Using the following thermodynamic identity,
\begin{equation}
 C_{V} = T \left(\frac{\partial S}{\partial T}\right)_{V},
\label{specific_heat_entropy}
\end{equation}
it is clear that the entropy $S$ is equal to the specific heat $C_{V}$ for the Free Electron model \cite{comment_1},
\begin{equation}
 S = \frac{N k_{B}\pi^{2}}{2} \frac{k_{B}T}{\epsilon_{F}} . 
\label{sommerfeld_entropy}
\end{equation}

Since we assume that the discharging process occurs at constant temperature, we can rewrite the Eq. (\ref{sommerfeld_entropy}) as 
\begin{equation}
 S(T,V,N) = A(T,V) N^{1/3}, 
\label{new_entropy}
\end{equation}
where 
\begin{equation}
 A(T,V) \equiv \frac{k_{B}^{2}\pi^{2}m}{\hbar^{2}} \left(\frac{V}{3\pi^{2}} \right)^{2/3}T > 0 .
\end{equation}

Now that we have the entropy as a function of all the relevant parameters of the problem, we are able to evaluate the entropy change of the system due to the discharge process. We denote the entropy of each capacitor $i$ in the initial $I$ or final $F$ configuration by by $S_I^{(i)}$ and $S_F^{(i)}$ respectively. The total
entropy change is given by the sum of the entropy changes of each capacitor,
\begin{equation}
 \Delta S = \Delta S_1 + \Delta S_2 = (S_F^{(1)} - S_I^{(1)}) + (S_F^{(2)} - S_I^{(2)}).
\label{entropy_change1}
\end{equation}

Before we go on performing the calculations, let's make an observation about the Free Electron Gas model. We assume that the piece of metal that we are modeling possess $N$ atoms, and that each one of these atoms contributes with $a$ electrons to our sample (typically, $a$ is $1$ or $2$). When an atom donates $a$ electrons, it will be positively charged, with $q = ae$, where $e$ is the electron's fundamental charge. So, when we consider all the sample, we conclude that our piece of metal is uncharged because the charge of the electrons cancels out the charge of the ions, although there exists free electrons. 

Furthermore, we are considering metal plates that could be positively or negatively charged. We know that the excess charge $q$ that a macroscopic metal can support is very small, when compared to the number of valence electrons in the sample. To see a good discussion on this subject, see \cite{feymann}, when Feynman introduces electric forces.

With these observations in mind, let us continue with our computations. Consider first the two plates of capacitor 1. Initially, one plate has an excess charge $q_{0}$, and the other, $-q_{0}$. We known that the electric charge is quantized, so we have $q_{0} = N_{0} e$, where $N_{0}$ is the excess number of electrons in the plate. Remembering that the total number of electrons in a plate is $Na$ plus the excess charge, we have

\begin{equation}
 \Delta S_1 = A(T,V)\left[\left(Na + \frac{q_{0}}{2e}\right)^{1/3} + \left(Na - \frac{q_{0}}{2e}\right)^{1/3} - \left(Na + \frac{q_{0}}{e}\right)^{1/3} - \left(Na - \frac{q_{0}}{e}\right)^{1/3}\right], 
\label{entropy_change_capacitor_1}
\end{equation}
\noindent
because in the equilibrium situation the excess charge in each plate is $\pm q_{0}/2$, as discussed before.

Analogously, for the capacitor 2, we have 

\begin{equation}
 \Delta S_2 = A(T,V)\left[\left(Na + \frac{q_{0}}{2e}\right)^{1/3} + \left(Na - \frac{q_{0}}{2e}\right)^{1/3} - 2\left(Na\right)^{1/3}\right]. 
\label{entropy_change_capacitor_2}
\end{equation}

\begin{figure}[!htb]
\begin{center}
\vspace{0.6cm}
\includegraphics[scale=0.8]{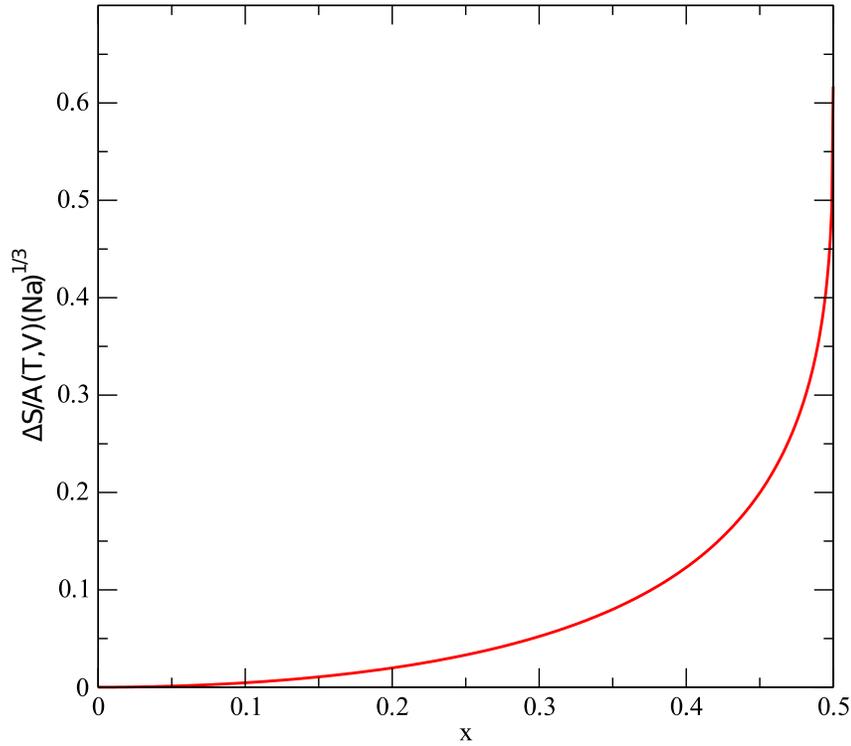}
\end{center}
\caption{Entropy change as a function of the ratio between the excess charge and the total number of valence electrons.}
\label{entropy_versus_x}
\end{figure}

Now, to obtain $\Delta S$, we substitute Eqs. (\ref{entropy_change_capacitor_1}) and (\ref{entropy_change_capacitor_2}) into (\ref{entropy_change1}): 

\begin{equation}
 \Delta S = A(T,V)(Na)^{1/3}\left[2\left(1 + x\right)^{1/3} + 2\left(1 - x\right)^{1/3} - \left(1 + 2x\right)^{1/3} - \left(1 - 2x\right)^{1/3} - 2\right] ,
\label{total_entropy_change}
\end{equation}
\noindent
where we defined $x \equiv q_{0}/2Nae$, the ratio between the excess charge and the number of valence electrons in the neutral sample. The expression obtained for $\Delta S$ in (\ref{total_entropy_change}) is always positive, as shown in Fig. (\ref{entropy_versus_x}). 

Although we obtained a expression for $\Delta S$, and have shown that $\Delta S > 0$, we can go a little further. As we discussed earlier, typically we have $x\ll 1$. Thereby, we can expand the Eq. (\ref{total_entropy_change}), using the binomial expansion for all terms. Doing this, and collecting only second order terms, we obtained

\begin{equation}
 \Delta S = \frac{4A(T,V)x^{2}(Na)^{1/3}}{9} = \frac{A(T,V)q_{0}^{2}(Na)^{-5/3}}{9e^{2}} .
 \label{simple_entropy_change}
\end{equation}
\noindent
In this case, we needed to consider second order terms in the expansion for $\Delta S$, because the first order term vanishes. This can be noticed visually in Fig.(\ref{entropy_versus_x}), as the slope of the curve vanishes as $x\rightarrow 0$.

\section{Energy Calculation}
 Up to now we have discussed the entropy change in the discharging process. We can also discuss what happens with the energy. First, we note a somewhat counter-intuitive fact. The electrostatic energy change $\Delta E_{El}$ of the system, between the initial and the final configurations does not depend on the resistance $R$. As long as $R>0$ \cite{comment_2}, we have 

\begin{equation}
\Delta E_{El} = -\frac{q_{0}^{2}}{4C} . 
\label{energy_change}
\end{equation}

However note that the eq.(\ref{energy_change}) only takes into account the electrostatic interaction. 
In order to obtain the total energy variation we should consider the energy change due to the
rearrangement of the energy levels of the free electron gas. At room temperature the total energy of the free electron gas is well approximated by its ground state energy given by
\begin{equation}
 E = \frac{3}{5}N\epsilon_F.
\label{free_elect_ene}
\end{equation}

Although in the entropic calculation we used the first order Sommerfeld expansion, in the energy calculation the use of the ground state energy is justified by the fact that in the latter $\Delta E \neq 0$ even for $T = 0\;$K. For the entropy, however, $\Delta S = 0$ for $T = 0\;$K and going beyond zeroth order is essential. 

\begin{figure}[!htb]
\begin{center}
\vspace{0.6cm}
\includegraphics[scale=0.5]{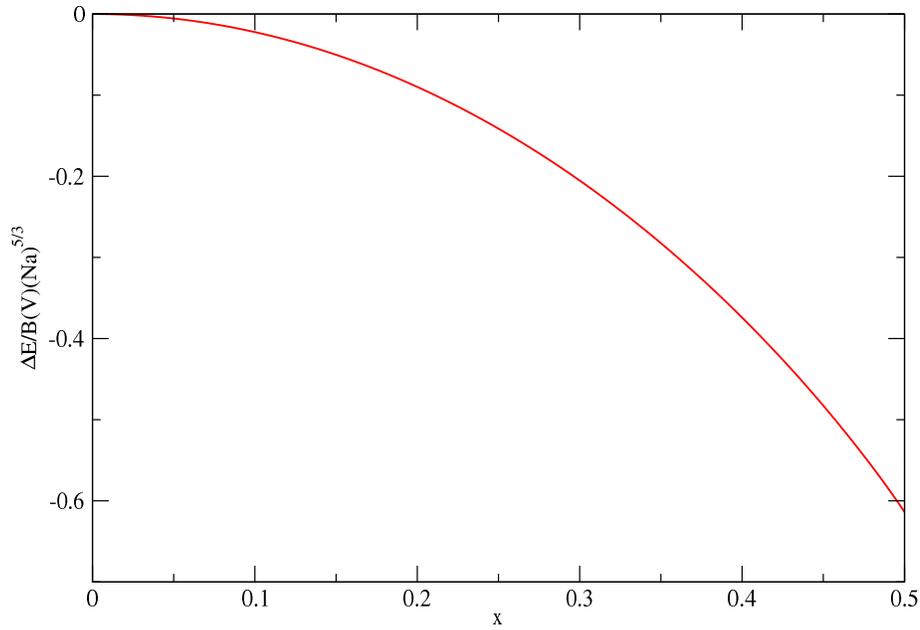}
\end{center}
\caption{Energy change as a function of the ratio between the excess charge and the total number of valence electrons.}
\label{energy_x}
\end{figure}

Using eq.(\ref{free_elect_ene}) to calculate the energy variation of the four capacitor plates, we obtain [see fig.(\ref{energy_x})], analogously to the entropy calculation,  
\begin{equation}
 \Delta E_{FE} = B(V) N^{5/3}\left[2 (1+x)^{5/3} + 2 (1-x)^{5/3}-(1+2x)^{5/3}-(1-2x)^{5/3}-2\right],
\label{deltaE_free_elect}
\end{equation}
 where
\begin{equation}
 B(V)=\frac{3}{5}\left(\frac{\hbar^2}{2m}\right)\left(\frac{3\pi^2}{V}\right)^{2/3}.
\end{equation}

Expanding (\ref{deltaE_free_elect}) for $x\ll 1$ we obtain
\begin{equation}
 \Delta E_{FE} = -\frac{20}{9}B(V)N^{5/3}x^2.
\end{equation}
  Thus, the total energy variation is given by
\begin{equation}
 \Delta E = -\frac{q_{0}^{2}}{4C}-\frac{20}{9}B(V)N^{5/3}x^2 .
\end{equation}

\section{Conclusions}

In this work we introduced a new approach for the two-capacitor problem. Instead of looking at it from 
an electromagnetic point of view, we focus on its thermodynamic properties, mainly the entropy change 
due to the discharge process. Our goal was to show that this process is irreversible, which means that 
the entropy variation must be always positive ($\Delta S > 0$). We assumed that the resistance is in 
thermal contact with a heat reservoir at the same temperature, which leads to  
$\Delta S_{R} + \Delta S_{HR} = 0$. Thus, all the entropy variation is due to the change in the 
electrons' distributions in the plates of the capacitors. We assumed that the valence electrons in the metal plates can be modeled by the Free Electron Gas Model. Under these considerations, we obtained $\Delta S > 0$ and $\Delta E<0$ for the whole process. 

\begin{acknowledgements}
 The authors are grateful for the financial support of the Brazilian agency Conselho Nacional de Desenvolvimento Cient\'ifico, CNPq, and to Nivaldo A. Lemos, Kaled Dechoum, Jorge S. S\'a Martins, Paulo Murilo C. de Oliveira, Marcio A. de Menezes and Marlon F. Ramos, for careful reading of the manuscript and helpful suggestions. 
\end{acknowledgements}




\vskip 2\baselineskip

\vskip 2\baselineskip


\begin{thebibliography}{40}
\bibitem{comment_2} The case $R = 0$ should be analyzed more carefully. See, for instance \cite{salih, boykin}.
\bibitem{comment_1} If you prefer a derivation that goes from the beggining, see \cite{fetter_walecka}.
\end{thebibliography}
\end{document}